\documentclass[10pt,conference]{IEEEtran}
\IEEEoverridecommandlockouts
\usepackage{cite}
\usepackage{amsmath,amssymb,amsfonts}
\usepackage{physics}
\usepackage{algorithmic}
\usepackage{graphicx}
\usepackage{textcomp}
\usepackage{xcolor}
\usepackage{commands}
\usepackage{acronym}
\usepackage[inline]{enumitem}
\usepackage{tabularx}
\usepackage{booktabs}
\usepackage{multirow}
\usepackage{makecell}
\usepackage{subcaption}
\usepackage{xspace}
\usepackage{siunitx}
\usepackage[most]{tcolorbox}
\usepackage{listings}
\usepackage{tikz}
\usetikzlibrary{decorations.pathreplacing,calc}
\usepackage[all]{nowidow}
\usepackage{hyperref}
\usepackage[switch]{lineno}

\acrodef{ai}[AI]{Artificial Intelligence}
\acrodef{ml}[ML]{Machine Learning}
\acrodef{llm}[LLM]{Large Language Model}
\acrodefplural{llm}[LLMs]{Large Language Models}
\acrodef{nl}[NL]{Natural Language}
\acrodef{nlp}[NLP]{Natural Language Processing}
\acrodef{vqe}[VQE]{Variational Quantum Eigensolver}
\acrodef{fci}[FCI]{Full Configuration Interaction}
\acrodef{qpe}[QPE]{Quantum Phase Estimation}
\acrodef{pf}[PF]{problem family}
\acrodefplural{pf}[PFs]{problem families}
\acrodef{tfim}[TFIM]{Transverse Field Ising Hamiltonian}
\acrodef{qubo}[QUBO]{Quadratic Unconstrained Binary Optimization}

\def\BibTeX{{\rm B\kern-.05em{\sc i\kern-.025em b}\kern-.08em
    T\kern-.1667em\lower.7ex\hbox{E}\kern-.125emX}}

\usepackage{todonotes}

\begin{document}

\title{Can LLMs Solve Science or Just Write Code? Evaluating Quantum Solver Generation}

\author{
\IEEEauthorblockN{Luciano Baresi}
\IEEEauthorblockA{\textit{Politecnico di Milano}\\
Milan, Italy\\
luciano.baresi@polimi.it}
\and
\IEEEauthorblockN{Domenico Bianculli}
\IEEEauthorblockA{\textit{University of Luxembourg}\\
Luxembourg, Luxembourg \\
domenico.bianculli@uni.lu}
\and
\IEEEauthorblockN{Maryse Ernzer}
\IEEEauthorblockA{\textit{University of Luxembourg}\\
Luxembourg, Luxembourg \\
maryse.ernzer@uni.lu}
\and
\IEEEauthorblockN{Livia Lestingi}
\IEEEauthorblockA{\textit{Politecnico di Milano}\\
Milan, Italy\\
livia.lestingi@polimi.it}
\and
\IEEEauthorblockN{Fabrizio Pastore}
\IEEEauthorblockA{\textit{University of Luxembourg}\\
Luxembourg, Luxembourg \\
fabrizio.pastore@uni.lu}
\and
\IEEEauthorblockN{Seung Yeob Shin}
\IEEEauthorblockA{\textit{University of Luxembourg}\\
Luxembourg, Luxembourg \\
seungyeob.shin@uni.lu}
}

\maketitle

\begin{abstract}
\acp{llm} show strong capabilities in code generation, motivating their use in automated quantum solver development. However, in quantum computing, successful execution of generated code is not sufficient: correctness depends on numerically accurate results, which are sensitive to non-trivial mappings, hybrid quantum-classical workflows, and algorithm-specific approximations. 
This work introduces \approachName, an iterative methodology to evaluate \acp{llm}' capability in generating quantum solvers for scientific
problems. 
The methodology adopts an iterative approach by executing the script generated by the \ac{llm}, comparing the result with the result of a classical solver, and refining the script until the two results match within a tolerance threshold. 
We empirically evaluated the methodology with five families of scientific problems of different complexities and five \acp{llm}, both open source and proprietary.
The results show that iterative refinement substantially improves success rates, but introduces a significant computational overhead.
Moreover, as model capability increases, failure modes shift from execution errors to numerical inaccuracies, highlighting the current limitations of \ac{llm}-based quantum software.
\end{abstract}


\section{Introduction}
\label{sec:intro}

Computational problems in physics, chemistry, and materials science require solving increasingly complex many-body models. 
In practice, these problems are predominantly addressed using classical numerical solvers, which constitute the standard toolchain in scientific computing~\cite{feynman2018simulating,cao2019quantum,mcardle2020quantum}.
However, with classical solutions, accurately simulating highly correlated quantum systems incurs computational costs that grow exponentially with system size, limiting scalability and making many realistic instances intractable~\cite{feynman2018simulating}.

Quantum computing has emerged as a promising alternative, offering algorithms that can more naturally represent and manipulate quantum states and potentially mitigate these scalability barriers~\cite{nielsen2010quantum, cao2019quantum}.
Despite the maturity of the underlying algorithms, translating them into correct and efficient quantum code remains a non-trivial and error-prone engineering task~\cite{Murillo-QCEfuture}, especially when they need to be tailored to a specific context (e.g., implementation with a specific language, integration into hybrid quantum-classical workflows).

As an example, computational physicists solve many closely related instances of a given problem, such as determining the properties of Hamiltonians that differ in particle number, interaction strength, encoding, or solver assumptions. 
Although textbook algorithms exist for these problems, adapting them to each concrete instance requires knowledge of programming languages, numerical libraries, and domain-specific APIs. 
Similarly, a quantum software engineer evaluating variational quantum algorithms may need to repeatedly implement and modify complex quantum code while avoiding subtle implementation errors that can invalidate numerical results. 
In both cases, maintaining a growing collection of hand-written scripts scales poorly and correctness is difficult to guarantee.

Recent advances in \acp{llm} have demonstrated strong capabilities in code generation and program synthesis~\cite{ravi2025llmloop}. 
This has motivated interest in using \acp{llm} to automate components of scientific computing workflows. 
However, existing work has largely focused on pedagogical use cases~\cite{robledo2023using,bitzenbauer2023chatgpt}, symbolic derivations~\cite{pan2025quantum}, or generating syntactically valid code~\cite{basit2025pennylang,campbell2025enhancing}. 
When tackling scientific computing with quantum algorithms,
this criterion is insufficient: a script may run successfully while producing inaccurate results, but existing studies assess \ac{llm} code generation capabilities
only by providing simple quantum circuit specifications, not complex scientific problems~\cite{guo2025quanbench,henderson2025programming}.

This work focuses on quantum computing approaches to scientific problems,
and investigates whether \acp{llm} can generate \emph{correct} solver code and under what conditions this is possible. 
Rather than proposing a new solver or algorithm, we introduce \approachName 
(Quantum-Solver Automated GEneration),
an execution-based methodology that treats quantum solvers as automatically generated artifacts whose correctness is numerically verified against a trusted reference (\ie a \emph{golden standard}). 
The methodology follows a generate-execute-verify loop, in which an \ac{llm} produces solver code, the code is executed, its output is compared against a ground truth within a specified tolerance, and feedback from execution is used to guide subsequent refinement.

\approachName addresses these challenges by combining \ac{llm}-based code generation with automatic execution and numerical verification. 
The methodology is not tied to a single scientific problem or solver: applying it to a different problem only requires the definition of a new prompt and a classical solver for verification. 
Our empirical evaluation focuses on five problem families that span different domains:
the Fermi-Hubbard model and the Transverse Field Ising Model from many-body physics~\cite{wecker2015solving, browaeys2020many, qin2022hubbard, stanisic2022observing}, the MaxCut problem from combinatorial optimization~\cite{lucas2014ising, peruzzo2014variational}, the Schwinger model from lattice gauge theory~\cite{martinez2016real, kokail2019self, nguyen2022digital, di2024quantum}, and molecular electronic structure problems from quantum chemistry~\cite{nam2020ground, lee2023evaluating}.

We evaluate \acp{llm} performance across the five problem families and different quantum algorithmic paradigms, including quantum-mapped solvers based on classical exact diagonalization, hybrid quantum-classical approaches such as the \ac{vqe}~\cite{peruzzo2014variational}, quantum annealing formulations, and circuit-based simulation techniques such as Trotterization and \ac{qpe}.
In addition, we evaluate \approachName~under two configurations that differ in the feedback
provided to the \acp{llm}
after the first iteration:
(1) only the output produced by the code generated in the previous iteration and 
(2) additionally the expected reference result from the classical solver.

The results reveal that \acp{llm} effectively generate correct solver code with iterative feedback and for problems with lower algorithmic complexity.
However, their performance degrades significantly for more complex settings, where failures are primarily due to numerical inaccuracies, rather than execution errors.
These findings highlight a capability limit for current \acp{llm}: 
iterative feedback improves success rates primarily within the first few iterations, but introduces significant overhead, while remaining failures are dominated by numerical inaccuracies.

In summary, our contributions are:
\begin{itemize}
    \item A methodology for evaluating \ac{llm}-generated scientific quantum code based on automatic execution and numerical verification.
    \item A benchmark for \acp{llm} on Hamiltonian solvers to assess end-to-end numerical correctness.
    \item An empirical comparison between quantum-mapped solvers using classical exact diagonalization, hybrid quantum-classical solvers based on \ac{vqe}, quantum annealing approaches, and circuit-based techniques such as Trotterization and \ac{qpe}.
\end{itemize}

The rest of the paper is structured as follows: 
\sref{rw} surveys the state of the art;
\sref{bg} outlines preliminary concepts underpinning our work;
\sref{meth} describes \approachName;
\sref{exp} presents experimental results;
\sref{concl} concludes the paper.

\section{State of the art}
\label{sec:rw}

The use of \acp{llm} for code generation and scientific problem solving has attracted significant attention in recent years. 
This section reviews previous work related to our study and clarifies how our approach extends the state of the art.

A growing body of work investigates \acp{llm} as tools for automatic code generation and program repair~\cite{LLM-code-survey}. 
Recent work has also investigated the use of \acp{llm} to support developers in understanding quantum programs, for example by generating and refining natural-language explanations of quantum algorithms~\cite{d2024exploring}.
General-purpose agentic frameworks such as LLMLOOP~\cite{ravi2025llmloop} employ iterative compile-execute-revise loops to improve generated code using execution feedback. These approaches typically evaluate success by syntactic correctness or by assessing whether the generated program executes without runtime errors. 
While effective for tasks such as code completion and bug fixing, where correctness can be adequately assessed through compilation success and passing test cases, such criteria are insufficient for scientific computing, where a program may run successfully without producing correct results, but test cases are unavailable, except if a reference implementation is provided.

Several studies explore the use of \acp{llm} in physics education and conceptual understanding~\cite{robledo2023using,bitzenbauer2023chatgpt}. These works assess \ac{llm} performance on explanatory tasks, problem-solving in natural language, or the generation of instructional material for undergraduate-level physics. While valuable in educational contexts, such studies do not address the generation of executable solver code for realistic scientific problems, nor do they evaluate numerical correctness against trusted references.

Recent work has also examined the ability of \acp{llm} to assist in symbolic derivations in quantum many-body physics, such as producing LaTeX expressions for Hamiltonians or deriving mean-field or Hartree-Fock equations~\cite{pan2025quantum}. These approaches operate purely at the symbolic level and do not attempt to generate or execute quantum solvers.

More closely related are efforts to benchmark \acp{llm} on quantum programming tasks. 
Frameworks such as PennyLang~\cite{basit2025pennylang} and QuanBench~\cite{guo2025quanbench} evaluate the ability of \acp{llm} to generate valid quantum circuits or code compatible with quantum software stacks such as Qiskit or PennyLane. Existing work assesses the correct generation of well known algorithms such teleportation~\cite{henderson2025programming}. These benchmarks primarily focus on circuit syntax, gate composition, or hardware-related considerations, and often evaluate correctness in terms of circuit validity or compilation success.

Our work focuses on end-to-end generation and verification of quantum solver code for realistic computational scientific problems.
Existing benchmarks assess correctness through test cases or compilation success, while
we evaluate whether \ac{llm}-generated quantum code produces numerically correct results, using a trusted classical solver. 
To our knowledge, this is the first study to evaluate \ac{llm}-generated quantum solver code through multi-turn execution-based numerical verification against trusted reference implementations.

\section{Background}
\label{sec:bg}

This section introduces key quantum computing concepts and technologies directly relevant to our work.
These include qubits, Hamiltonian representations, and the main classes of quantum algorithms considered.

In contrast to  classical computations, quantum computing arises by replacing a classical bit that can be either in the state ``0'' or ``1'', with a quantum bit (qubit) that can exist in the complex superposition of its basis states $\ket{0}$ and $\ket{1}$ such as $\psi = \alpha \ket{0} + \beta \ket{1} $ with the complex parameters $\alpha$ and $\beta$.

The unique properties of quantum systems resulting from the superposition of states and strong correlations, such as quantum entanglement, allow for parallel and distributed processing, and come with an expected increase in computational efficiency.
At the same time, these properties translate into specific challenges in the code generation related to quantum computing such as handling of coherence and measurement in circuit constructions, and qubit-operator mappings.
Indeed, to simulate a given system, its degrees of freedom have to be mapped onto the qubit framework and restricted to the appropriate symmetry sectors, which in most cases require a nontrivial mapping such as Jordan-Wigner, parity, and Bravyi-Kitaev transformations~\cite{tranter2018comparison, seeley2012bravyi, bravyi2002fermionic}. 

Among the fundamental building blocks of quantum algorithms that have sparked the idea of an advantageous computational scaling for gate-based quantum computers is the Quantum Phase Estimation (QPE), used to determine the ground-state energy via the phase of the eigenvalue~\cite{kitaev1995quantum, fomichev2024initial}.
More recently, hybrid quantum-classical variational algorithms, such as the variational quantum eigensolver (VQE), have been developed to enable eigenvalue estimation considering constraints in terms of feasible circuit depth and gate fidelity imposed by near-term hardware~\cite{peruzzo2014variational}. 

Adding to the toolbox of approaches to find the lowest eigenvalue of certain specific systems, quantum annealing has emerged as a near-time alternative based on the quantum adiabatic algorithm~\cite{hauke2020perspectives}.

In addition to the eigenvalue problem, the simulation of many-body Hamiltonian evolution is one of the originally proposed useful applications of quantum simulators~\cite{feynman2018simulating, nielsen2010quantum}.
On a discrete quantum computer, Hamiltonian time-dynamics can be simulated via the basic Trotterization algorithm that repeats time-evolution for a small time-step~\cite{lloyd1996universal, clinton2021hamiltonian}.

These algorithms are universally applicable to problems that can be cast in Hamiltonian form and thus span an extensive range of computational tasks found in natural sciences from quantum chemistry, over many body-physics and even quantum field theory~\cite{sakamoto2024end}.

\section{The \approachName Methodology}
\label{sec:meth}

\begin{figure}[t]
\centering
    \includegraphics[width=\linewidth]{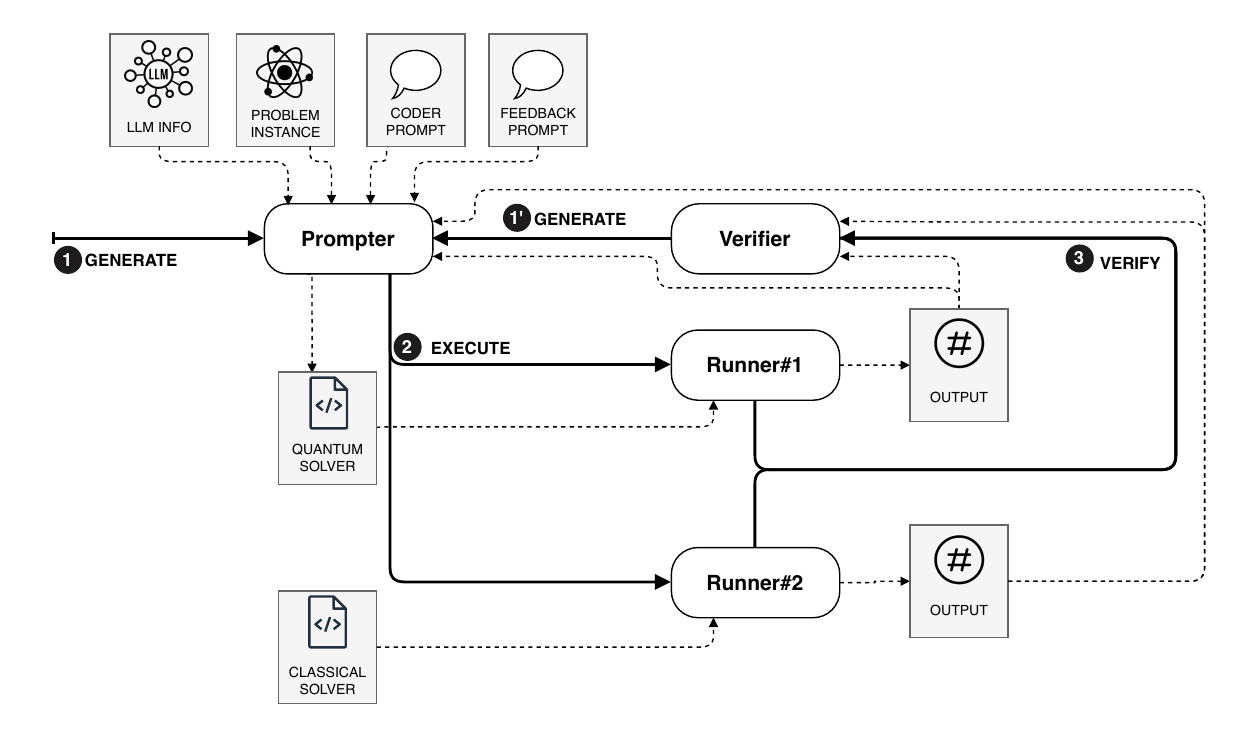}
    \caption{\approachName{} workflow. Solid arrows represent operational steps numbered according to their execution while dashed arrows represent input/output artifacts.}
    \label{fig:fw}
\end{figure}

\begin{figure}[t]
\centering
    \begin{subfigure}{\columnwidth}
    \centering
        \includegraphics[width=\columnwidth]{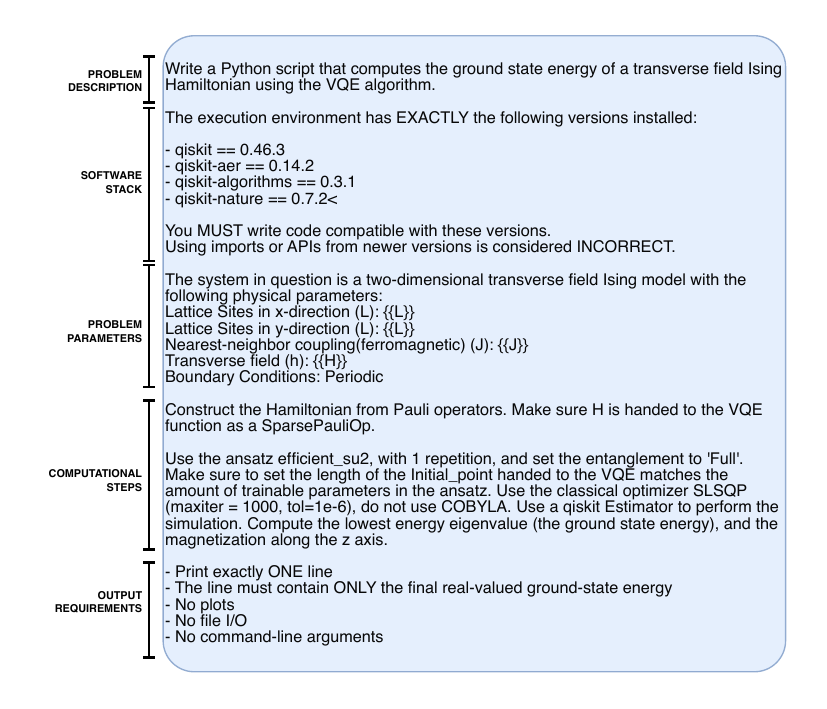}\vspace{-1em}
        \caption{Coder prompt.}
        \label{fig:coderprompt}
    \end{subfigure}
    \begin{subfigure}{\columnwidth}
    \centering
        \includegraphics[width=\columnwidth]{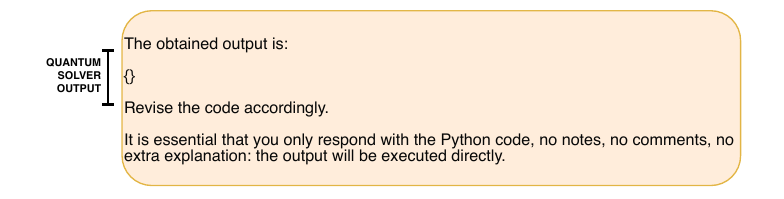}\vspace{-0.2em}
        \caption{Feedback prompt.}
        \label{fig:feedbackprompt}
    \end{subfigure}
    \begin{subfigure}{\columnwidth}
    \centering
        \includegraphics[width=\columnwidth]{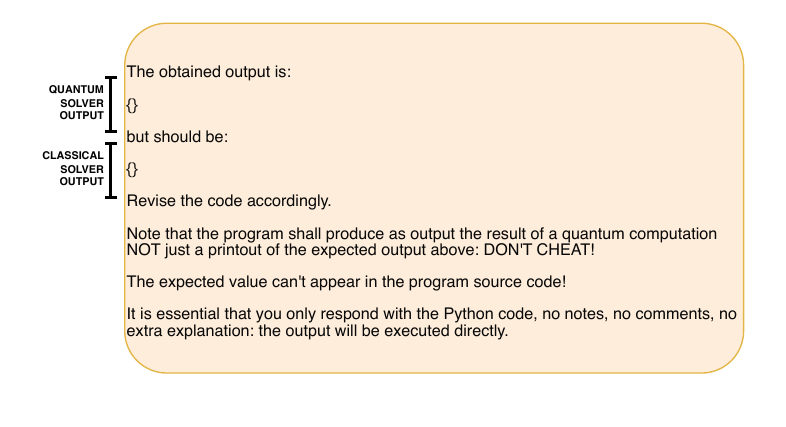}\vspace{-1.5em}
        \caption{Informed feedback prompt.}
        \label{fig:inffeedbackprompt}
    \end{subfigure}
\caption{Prompt templates.}
\label{fig:prompts}
\end{figure}

This section presents \approachName, an execution-based methodology for evaluating the ability of \acp{llm} to generate correct quantum solver code.
\approachName combines automated code generation, controlled execution, and numerical verification in a multi-turn iterative loop (represented in \fref{fw}), allowing the assessment of both functional correctness and failure behavior under realistic conditions.

In scientific computing, correctness cannot be inferred from execution success alone, nor from discrete test cases. 
Therefore,
\approachName is designed around three principles: 
\begin{enumerate*}[label=(\arabic*)]
    \item correctness is defined in terms of numerical agreement with a trusted reference, 
    \item failures are detected through execution-based numerical verification, which is infeasible via static analysis, and 
    \item refinement is guided by execution feedback in a closed loop.
\end{enumerate*}

\subsection{Execution-Based Feedback Loop}

In scientific computing,
a trusted reference result cannot be obtained by reading the problem specifications but requires the execution of a reference mathematical simulation, which might be unavailable. Engineers may rely on \acp{llm} to produce quantum code without having a reference implementation: it is therefore important to assess how likely \acp{llm} produce numerically correct results. 

In \approachName, given a problem instance, an \ac{llm} is prompted to generate a quantum solver implementation (step~\circnum{1}). The generated code is executed in a controlled environment (step~\circnum{2}), and its output is compared with a reference value provided by a classical solver (step~\circnum{3}).
If the output matches the reference within a predefined tolerance, the solution is considered correct. Otherwise, if the generated code fails to execute correctly or produce an output within tolerance, execution feedback is provided to the \ac{llm} and a new iteration is initiated (step~\circnum{1'}).
The generate-execute-verify-iterate loop (shown in \fref{fw}) continues until a correct solution is found or a predefined iteration budget is exhausted.

\subsection{Problem Definition}

The structure of the execution-based loop methodology is designed to be problem-agnostic. No problem-specific logic is embedded in the methodology itself.
The target scientific problem is characterized by means of artifacts that the methodology receives as inputs, specifically:
\begin{enumerate}[label=(\roman*)]
    \item a machine-readable description of the problem,
    \item the \emph{coder} prompt initially describing the problem for the first iteration,
    \item the \emph{feedback} prompt for all subsequent iterations, and
    \item a classical solver that computes the reference result.
\end{enumerate}
As a result, changing the target problem---such as switching from one physical model to another, or from a classical to a hybrid quantum algorithm---requires modifying said artifacts, without altering the surrounding infrastructure.

The machine-readable description of the problem instance contains a descriptor for the problem (\eg ``condensedmatter/tfim''), the selected set of parameters (\eg lattice sites, nearest-neighbor coupling, and transverse field) and the parameter values characterizing the problem instances of interest.

The coder prompt (of which \fref{coderprompt} reports an example) specifies the problem and the conditions under which the solution must be generated. Crafting the prompt requires knowledge of the domain, the implementation constraints, and the output requirements to guide the \ac{llm} toward generating a meaningful solver.
In more detail, the prompt defines the required software stack (\eg specific library versions and APIs), ensuring that the generated code is compatible with the controlled environment where it will run.
The prompt also describes the scientific problem (\eg Hamiltonian definition, boundary conditions, and physical symmetries) and enforces correct representations (such as fermion-to-qubit mappings).
This portion may be customized on the basis of the parameter values of the specific problem instance. 
Depending on the complexity of the problem, it might be necessary to include a rough outline of the required computational steps or, in the most complex cases, precise step-by-step instructions.
The prompt also restricts the expected format of the output (\eg numeric values only) for automated parsing and, finally, diagnostic hooks for downstream verification and debugging.

At each iteration, the \ac{llm} is informed of the entire conversation history. Therefore, the feedback prompt does not contain any information on the problem, but only informs the \ac{llm} about the outcome of the previous iteration, instructing it to fix the code accordingly. 
Depending on the template used for the feedback prompt, it is possible to only provide the \ac{llm} with the output of the quantum solver generated in the last iteration (\eg the exception that was raised or the incorrect numeric output produced) or additionally include the output of the classical solver (\ie the expected output).
Figures~\ref{fig:feedbackprompt} and \ref{fig:inffeedbackprompt} show the two templates with and without the expected output, respectively.

The classical solver is a domain-specific numerical implementation that computes a reference solution for a problem instance. 
For the problems selected in this work, many-body and spin systems are addressed through explicit or matrix-free Hamiltonian constructions combined with exact or iterative diagonalization (\eg sparse eigensolvers or Lanczos methods). 
Combinatorial problems are mapped to diagonal Ising Hamiltonians and solved via exact diagonalization with brute-force verification. 
Dynamical models rely on full Hamiltonian construction and time evolution in constrained bases. Quantum chemistry instances are treated using \emph{ab initio} methods followed by full configuration interaction.
In summary, these solvers rely on exact or systematically convergent methods to ensure high-fidelity reference values in tractable regimes.

\subsection{Software Components}

\approachName architecture consists of three main components, which implement the execution-based evaluation loop and whose roles are illustrated in \fref{fw}.

The prompter is responsible for interacting with the \ac{llm}.
It loads a prompt template describing the target task, submits it to the model, and records the generated code. In subsequent iterations, the prompter incorporates execution feedback into the prompt to guide refinement. The prompter also maintains a record of the full interaction history for reproducibility and analysis.
The technical information required to access the \ac{llm}, such as authentication tokens, API endpoints, and model identifiers, is also provided as input to the prompter.

The runner executes solver code in a controlled environment. Standard output and error streams are captured and logged. Execution failures, runtime errors, and numerical output are all recorded and made available to downstream components. 
As per \fref{fw}, two instances of the runner are necessary: one for the generated quantum solver and one for the classical solver that runs concurrently.
Proceeding to the verification step requires both executions to have terminated. Therefore, the runner also enforces timeouts (which are properly tuned for the problem at hand with preliminary analysis) and resource limits to prevent pathological executions.

Concerning the verification step, we distinguish between three related concepts to avoid ambiguity.
The classical solver described previously is a numerical implementation that computes a reference result for a problem instance.
The \emph{verifier} is the component responsible for comparing the output of the generated quantum solver with the reference value computed by the classical solver. To this end, the solver applies a numerical tolerance (whose tuning requires domain knowledge) and produces a pass/fail result.
We refer to the combination of these two elements as the \emph{oracle}, which defines the correctness criterion adopted by \approachName.

All outputs from each iteration, including generated code, execution logs, and verification results, are persistently recorded, resulting in a repository of test episodes that serves as a basis for analysis, comparison, and reproducibility.

\section{Evaluation}
\label{sec:exp}

Our evaluation addresses the following research questions:
\begin{enumerate}[label=\textbf{RQ\arabic*.},leftmargin=*]
    \item \emph{How effectively can state-of-the-art \acp{llm} generate correct quantum code for computational scientific problems?}
\end{enumerate}
From a practitioner perspective, it is essential to know whether \acp{llm} can reliably generate correct quantum solver implementations for scientific problems. This question informs software engineers and physicists about the extent and conditions under which \acp{llm} can reduce the effort to manually code and produce numerically valid results.
\begin{enumerate}[label=\textbf{RQ\arabic*.},leftmargin=*]
\setcounter{enumi}{1}
    \item \emph{When the \ac{llm} fails to produce correct code, what are the most common causes of failures?}
\end{enumerate}
When \ac{llm}-generated code fails, understanding the underlying causes is critical to diagnosing problems and improving workflows. This question provides actionable information for software engineers on the current limitations of \acp{llm} and for physicists on the types of errors that can compromise the validity of generated solutions.
\begin{enumerate}[label=\textbf{RQ\arabic*.},leftmargin=*]
\setcounter{enumi}{2}
    \item \emph{What is the overhead induced by \approachName?}\end{enumerate}
Even when \acp{llm} produce correct solutions, their practical usefulness also depends on the computational cost required to achieve this. This question informs practitioners on the trade-off between iterative refinement and execution time, helping them assess whether the approach is viable in their applications.

\subsection{Design of the Experiments}

\subsubsection{Evaluation Subjects}
\label{sec:subjects}

\begin{table}[t]
\centering
\caption{Evaluation subjects.}
\label{tab:subjects}
\resizebox{\columnwidth}{!}{
    \begin{tabular}{l l l l l}
        \toprule
        \textbf{PFID} & \textbf{Problem Family} & \textbf{Domain} & \textbf{\makecell[l]{Quantum\\Algorithm}} & \textbf{Parameters} \\ 
        \midrule
        \multirow{3}{*}{\pf{1}} & \multirow{3}{*}{Fermi Hubbard} & \multirow{3}{*}{Condensed Matter} & \multirow{3}{*}{VQE} & Lattice size ($L$) \\
        & & & & Hopping strength ($t$) \\
        & & & & On-site Interaction ($U$) \\[.5em]
        \multirow{3}{*}{\pf{2}} & \multirow{3}{*}{\makecell[l]{Transverse Field\\Ising Hamiltonian}} & \multirow{3}{*}{\makecell[l]{Many-body\\Statistical Physics}} & \multirow{3}{*}{VQE} & Lattice size ($L$) \\
        & & & & Coupling ($J$) \\
        & & & & Transverse field ($h$) \\[.5em]
        \pf{3} & MaxCut & \makecell[l]{Combinatorial\\Optimization} & \makecell[l]{Quantum\\Annealing} & Weighted Graph ($N, E$) \\[.5em]
        \multirow{3}{*}{\pf{4}} & \multirow{3}{*}{Schwinger Model} & \multirow{3}{*}{Lattice Gauge Theory} & \multirow{3}{*}{Trotterization} &  Lattice size ($L$) \\
        & & & & Hopping strength ($h$) \\
        & & & & Gauge coupling ($g$) \\[.5em]
\pf{5} & \makecell[l]{Molecular Electronic\\Structure} & Quantum Chemistry & QPE & Bond Length ($\mathit{BL}$)\\
        \bottomrule
    \end{tabular}
}
\end{table}

The evaluation targets five \acp{pf} summarized in \tref{subjects}. 
The selected problems range from solid state physics, combinatorics, quantum chemistry, and quantum field theory. They
possess parameter regions rich in interesting phenomena and challenging to explore using solely classical methods.
The evaluation comprises four instances of each \ac{pf}, each with a different set of parameter values, which are available in our artifact repository. 
For verification purposes, in this work, we restrict the parameter range to values for which the classical solvers can compute a solution within a budget. 
Classical solvers used as reference scale exponentially with system size, making larger instances computationally too intensive. In concrete terms, the budget is defined by the available computational resources and a timeout limit per-run, which varies between \acp{pf}.

Within the field of quantum physics, we study three different problems with different levels of interaction complexity. In \pf{1}, we consider
fermionic particle statistics and hopping between lattice sites, 
with the associated Hamiltonian given by the Fermi-Hubbard model. 
The model describes instance systems of strongly correlated electrons commonly encountered in solid-state physics and can be used to study phenomena such as high temperature superconductivity~\cite{keimer2015quantum}.
We explore the regions of strong interactions ($U/t \gg 1$) and solve for the ground-state using the \ac{vqe} algorithm.
\pf{1} constitutes the most complex problem among our subjects requiring a timeout of \SI{3000}{\s}, while \SI{300}{\s} are sufficient in all other four cases. 

\pf{2} focuses on
the simple case of localized spin interactions captured by the \ac{tfim} model of spin-exchange dynamics with nearest-neighbor interactions and a transverse magnetic field. This model can, for example, be used to study quantum phase transitions in magnetic systems~\cite{heyl2013dynamical}. We solve the Hamiltonian for its energetic ground-state by applying the \ac{vqe} algorithm.
A simplified variant of this model, the Ising model, can be used to solve combinatorial problems such as the MaxCut problem (\pf{3}) belonging to the \ac{qubo} problem family. It consists of finding a subgraph with a maximal amount of edges of a planar graph, mapped to the Ising Hamiltonian, solved for its ground-state using quantum annealing~\cite{lucas2014ising, kadowaki1998quantum}. The lowest energy value corresponds to the optimized solution of the associated combinatorial problem. 

Adding relativistic fermionic fields with gauge interactions to the picture results in the 1-dimensional massive Schwinger Hamiltonian employed as a toy model for lattice gauge theories such as quantum electrodynamics to capture phenomena such as spontaneous symmetry breaking and particle-antiparticle interactions~\cite{martinez2016real}. We evaluate its dynamical evolution using a trotterization algorithm (\pf{4}). 

Finally, we address the common quantum chemistry problem (\pf{5}) of finding the ground-state of a molecular electronic structure with the example of a hydrogen molecule with \ac{fci} and different bond length values and solve it by \ac{qpe}~\cite{aspuru2005simulated}.

\subsubsection{Methods Under Comparison}

\begin{table}[t]
\centering
\caption{\acp{llm} selected for the evaluation.}
\label{tab:llms}
\resizebox{\columnwidth}{!}{
    \begin{tabular}{l l l l l}
        \toprule
        \textbf{ID} & \textbf{Model} & \textbf{Vendor} & \textbf{Date} & \textbf{Params} \\
        \midrule
        DS-3.2 & \href{https://huggingface.co/deepseek-ai/DeepSeek-V3.2}{DeepSeek-V3.2} & DeepSeek & Dec 1, 2025 & 685B \\
        GPT-OSS & \href{https://huggingface.co/openai/gpt-oss-120b}{gpt-oss-120b} & OpenAI & Aug 8, 2025 & 120B \\
        QWEN-480B & \href{https://huggingface.co/Qwen/Qwen3-Coder-480B-A35B-Instruct}{Qwen3-Coder-480B-A35B-Instruct} & Alibaba Cloud & Aug 22, 2025 & 480B\\
        QWEN-FP8 & \href{https://huggingface.co/Qwen/Qwen3-Coder-Next-FP8}{Qwen3-Coder-Next-FP8} & Alibaba Cloud & Feb 3, 2026 & 80B \\
        GPT-5.2 & \href{https://developers.openai.com/api/docs/models/gpt-5.2}{gpt-5.2} & OpenAI & Dec 11, 2025 & 2-50T \\
        \bottomrule
    \end{tabular}
}
\end{table}

We compare \approachName
with a single-turn approach serving as the baseline method. 
The latter is the first iteration of \approachName, with the \ac{llm} prompted once without further refinement.
This isolates the effect of iterative feedback and refinement in subsequent turns.

Moreover, we evaluate two configurations of \approachName that differ in the feedback prompt template. In the first configuration, from the second iteration forward, the \ac{llm} under test is only informed of the output that its previous response produced (shown in \fref{feedbackprompt}). 
In the second configuration, the \ac{llm} is additionally informed of the expected numerical output computed through the verifier script (shown in \fref{inffeedbackprompt}).

Each method is applied to five different \acp{llm} specialized in text generation, covering different parameter sizes and vendors, and including open source and proprietary models.
The characteristics of the selected  \acp{llm} are summarized in~\tref{llms}.

\subsubsection{Metrics and Statistical Tests}

We assess the different approaches in terms of \emph{success rate} versus \emph{time overhead}. More precisely, let ${I\in\mathbb{N}}$ be the number of instances per \ac{pf}, ${R\in\mathbb{N}}$ the number of repetitions per \ac{pf} instance, and ${T\in\mathbb{N}}$ the budget of turns per repetition.
Variable ${X_{j,k}^{(m,\mathit{pf})}\in[1, T]}$ denotes the turn at which model $m$ succeeds on $\mathit{pf}$ instance $j$ during repetition $k$.
We then assess the \acp{llm} effectiveness through $\mathbf{1}(X_{j,k}^{(m,\mathit{pf})}\leq t)$ indicating whether success occurs by turn $t$, referred to as \success{t}.
Given the stochastic nature of \ac{llm} text generation, experiments are performed with four instances for each \ac{pf} (${I=4}$), 10 runs (${R=10}$), and a budget of 10 turns per run (${T=10}$).

The evaluation follows the guidelines by Arcuri and Briand~\cite{arcuribriand}.
We assess the statistical significance of the differences between \approachName~and the single-turn baseline, both in terms of \success{t} and execution time, using the Mann-Whitney U test, and we report the Vargha-Delaney effect size measure to characterize the size of these differences~\cite{vargha}.
We adopt the standard interpretation of effect sizes, reporting small, medium, and large effects for values greater than $0.55$, $0.63$, and $0.70$, respectively.

\subsubsection{Evaluation Testbed}

All experiments are performed on a commodity machine running Ubuntu 24.04, equipped with 48 CPUs, 64~GB of memory, and a base clock speed of 2.20~GHz.
All \acp{llm} are accessed through the Hugging Face\footnote{\href{https://huggingface.co}{https://huggingface.co}.} API except for gpt-5.2 for which we rely on the proprietary API.

\subsection{Results}

\subsubsection{RQ1 (Effectiveness)}

\begin{figure*}[t]
\centering
    \begin{subfigure}{2\columnwidth}
    \centering
        \includegraphics[width=.3\columnwidth]{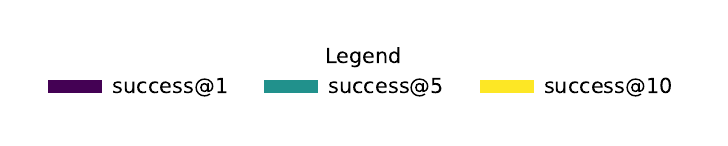}
    \end{subfigure}
    \begin{subfigure}{.95\columnwidth}
        \centering
        \includegraphics[width=\columnwidth]{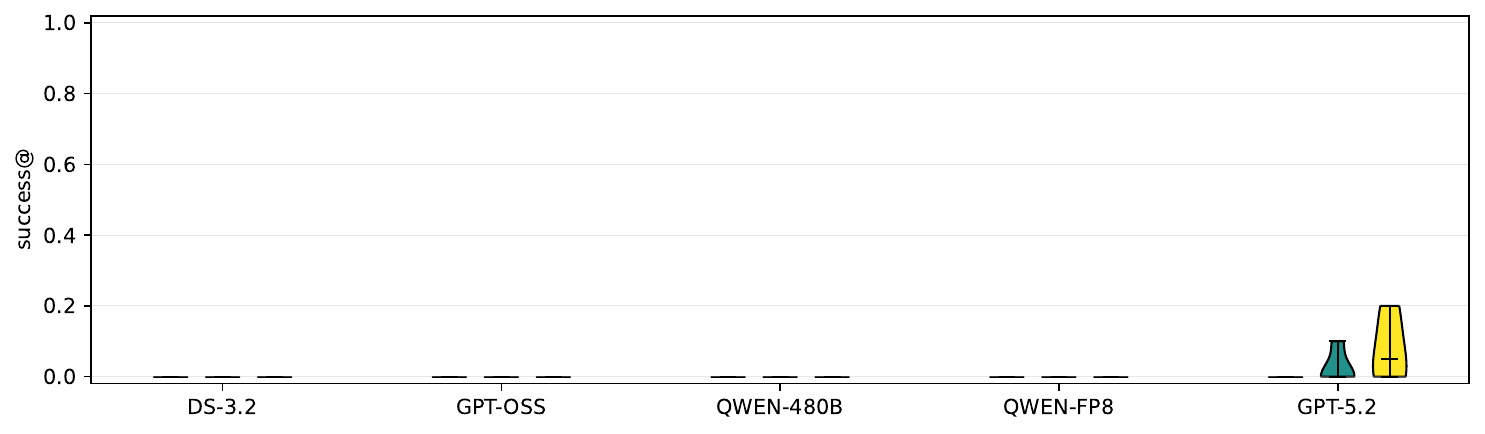}
        \caption{PF1 with feedback prompt.}
        \label{fig:rq1pf1}
    \end{subfigure}
    \begin{subfigure}{.95\columnwidth}
        \centering
        \includegraphics[width=\columnwidth]{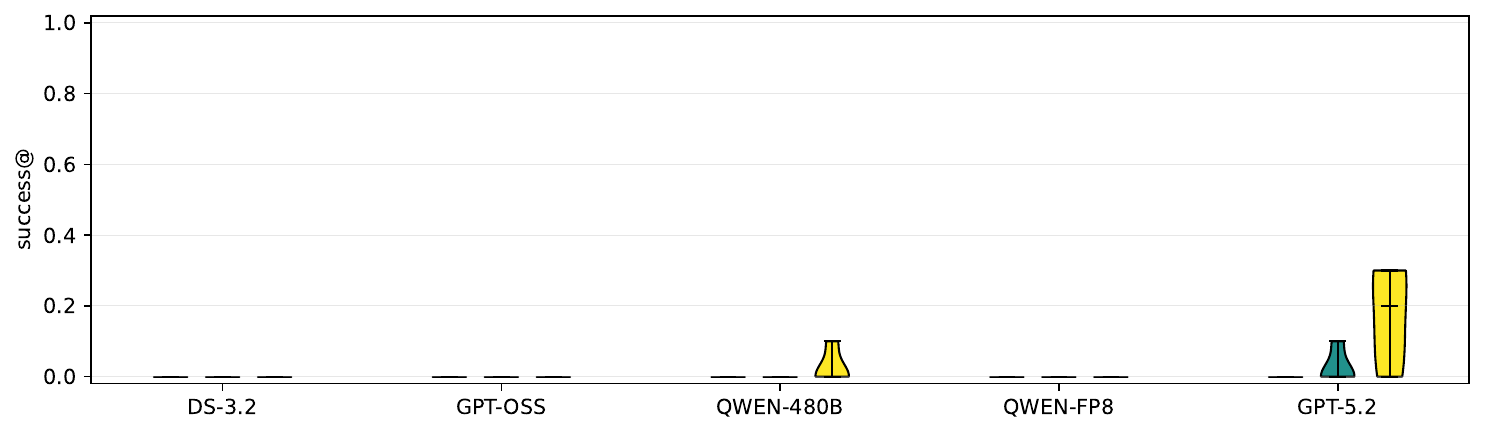}
        \caption{PF1 with informed feedback prompt.}
        \label{fig:rq1pf1inf}
    \end{subfigure}
    \begin{subfigure}{.95\columnwidth}
        \centering
        \includegraphics[width=\columnwidth]{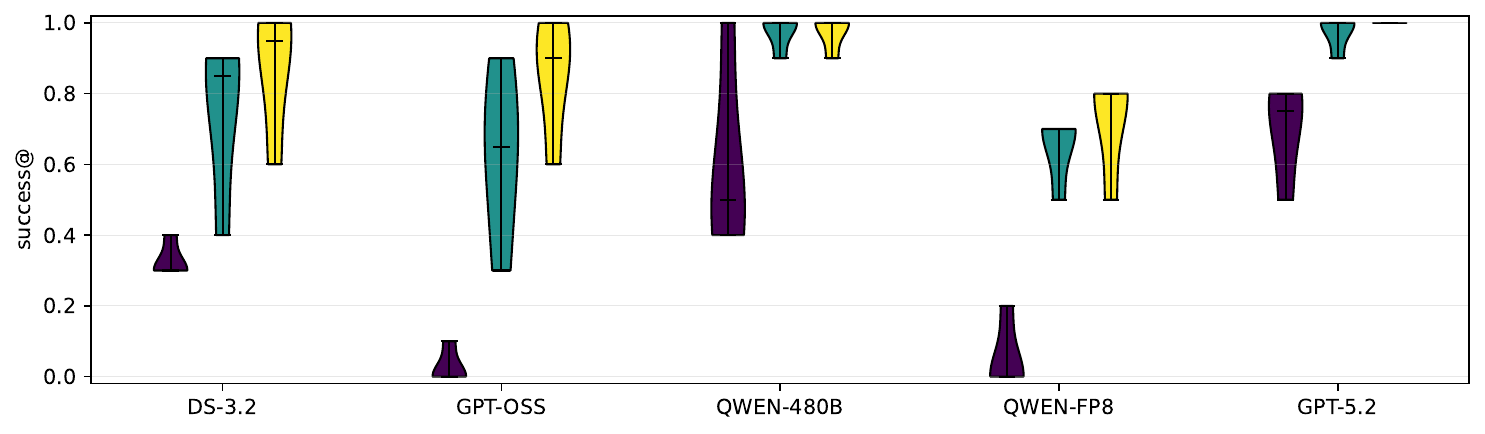}
        \caption{PF2 with feedback prompt.}
        \label{fig:rq1pf2}
    \end{subfigure}
    \begin{subfigure}{.95\columnwidth}
        \centering
        \includegraphics[width=\columnwidth]{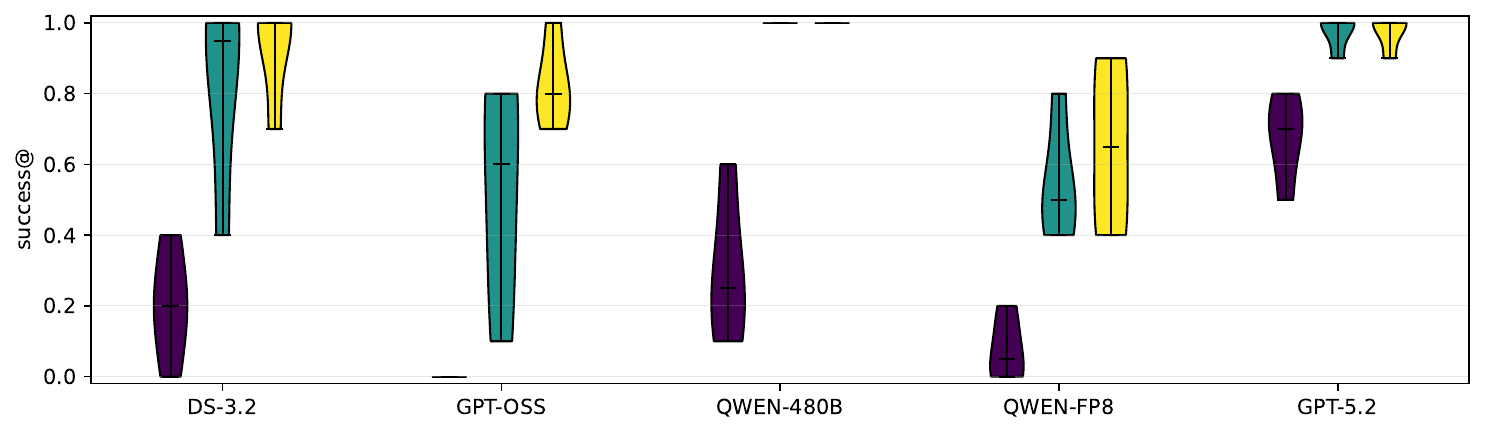}
        \caption{PF2 with informed feedback prompt.}
        \label{fig:rq1pf2inf}
    \end{subfigure}
    \begin{subfigure}{.95\columnwidth}
        \centering
        \includegraphics[width=\columnwidth]{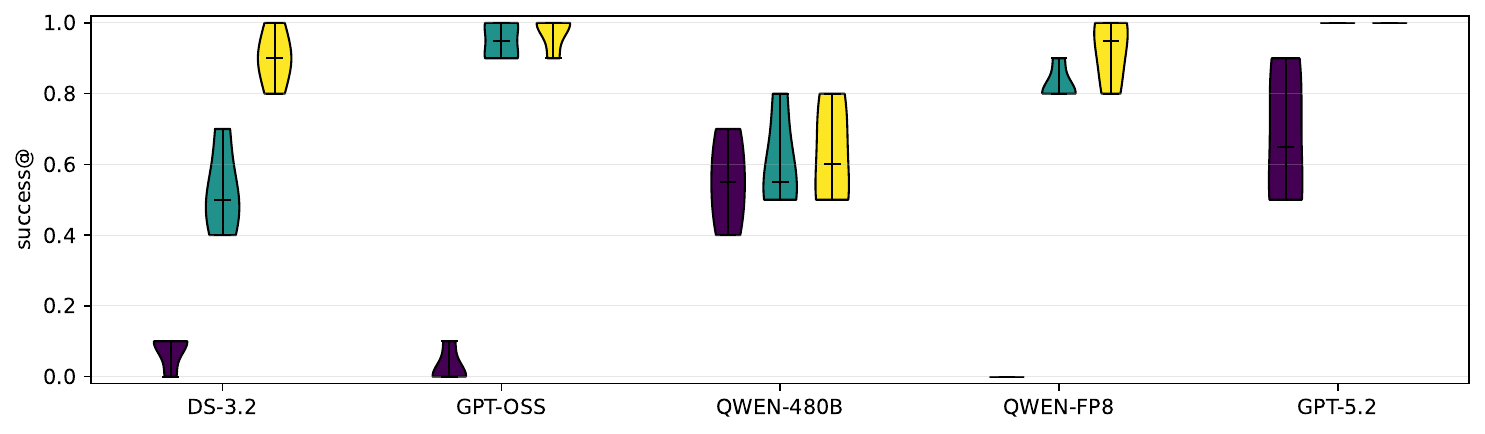}
        \caption{PF3 with feedback prompt.}
        \label{fig:rq1pf3}
    \end{subfigure}
    \begin{subfigure}{.95\columnwidth}
        \centering
        \includegraphics[width=\columnwidth]{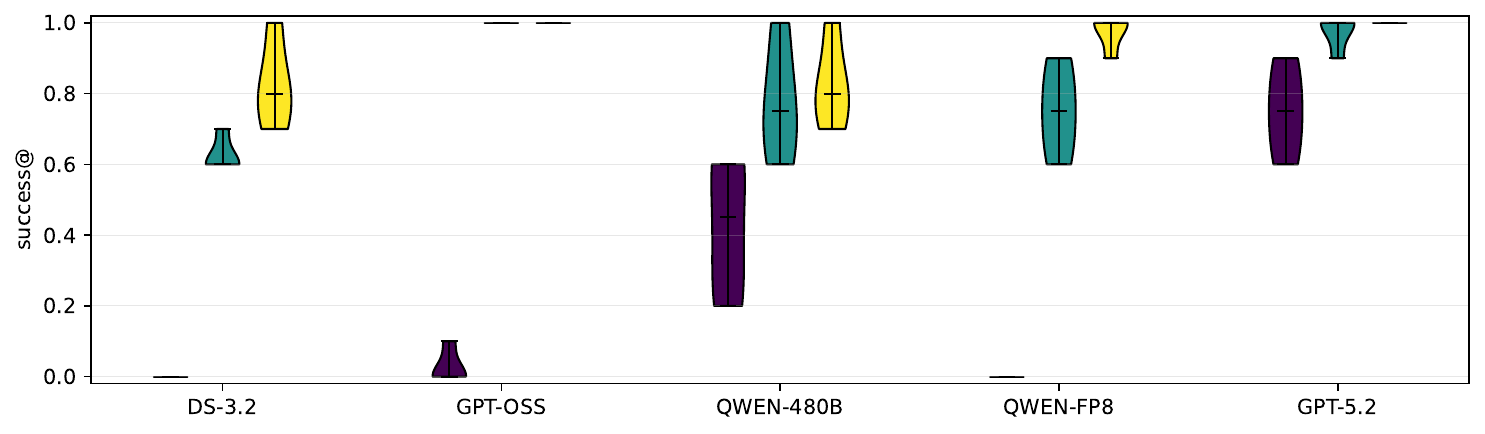}
        \caption{PF3 with informed feedback prompt.}
        \label{fig:rq1pf3inf}
    \end{subfigure}
    \begin{subfigure}{.95\columnwidth}
        \centering
        \includegraphics[width=\columnwidth]{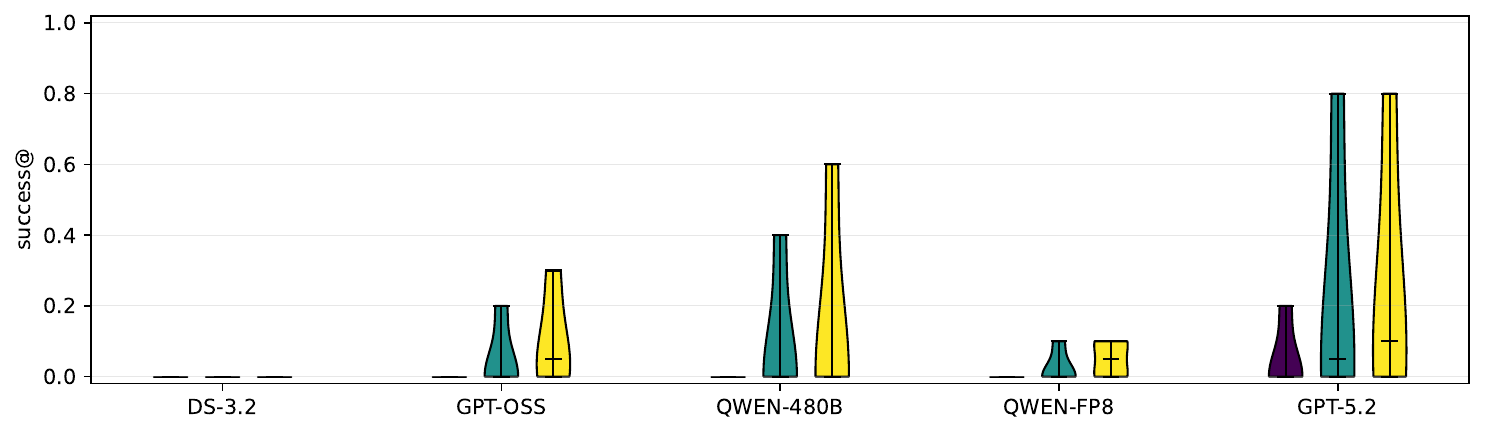}
        \caption{PF4 with feedback prompt.}
        \label{fig:rq1pf4}
    \end{subfigure}
    \begin{subfigure}{.95\columnwidth}
        \centering
        \includegraphics[width=\columnwidth]{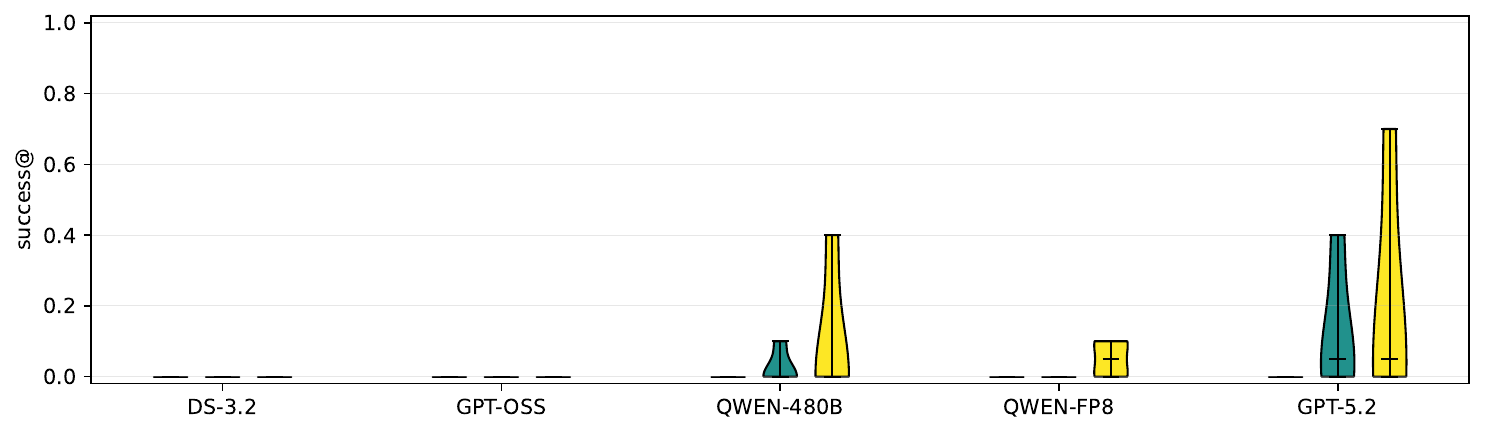}
        \caption{PF4 with informed feedback prompt.}
        \label{fig:rq1pf4inf}
    \end{subfigure}
    \begin{subfigure}{.95\columnwidth}
        \centering
        \includegraphics[width=\columnwidth]{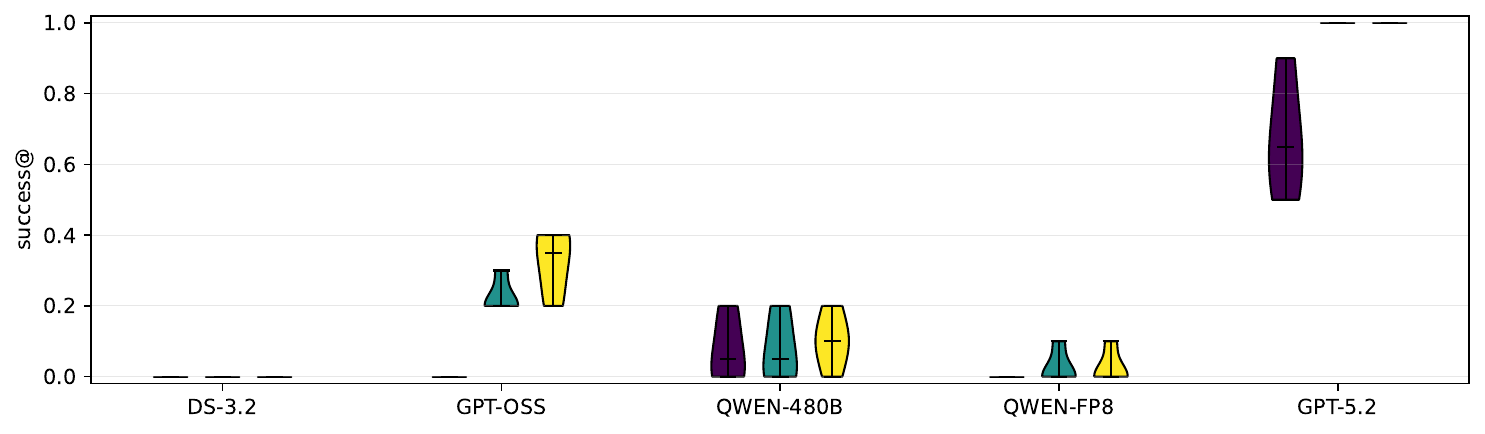}
        \caption{PF5 with feedback prompt.}  
        \label{fig:rq1pf5}
    \end{subfigure}
    \begin{subfigure}{.95\columnwidth}
        \centering
        \includegraphics[width=\columnwidth]{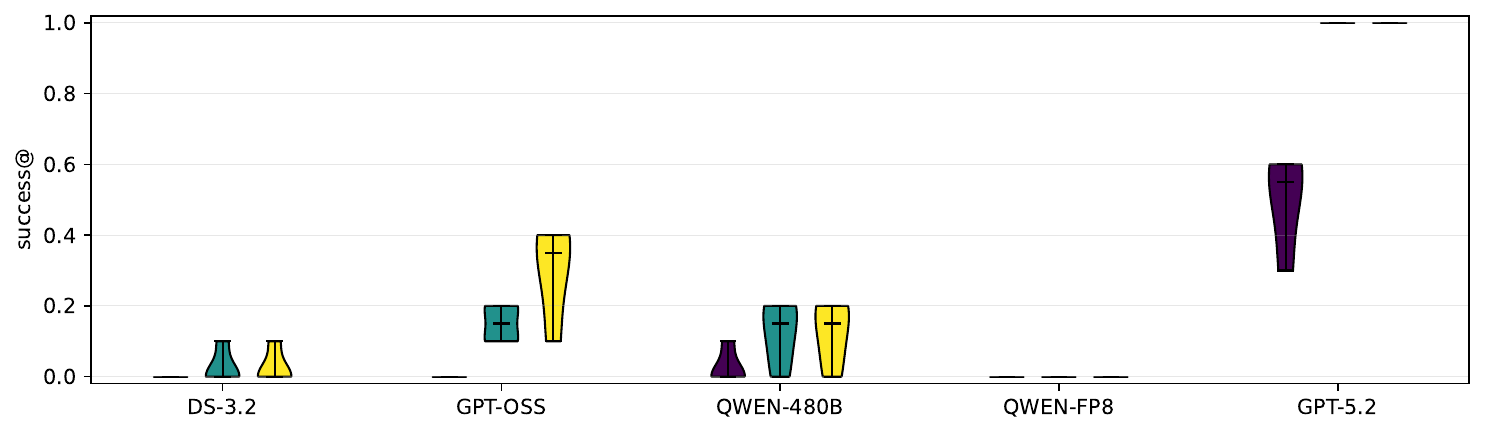}
        \caption{PF5 with informed feedback prompt.}
        \label{fig:rq1pf5inf}
    \end{subfigure}
\caption{RQ1 results showing \success{1}, \success{5}, and \success{10} distributions by \ac{pf} and feedback prompt template.}
\label{fig:rq1}
\end{figure*}

\begin{table}[t]
\centering
\caption{Statistical comparison of \success{} distributions split by \ac{pf}, feedback prompt variant, and model. Each cell reports the p-value of the Mann--Whitney U test and Vargha--Delaney effect size category (N, S, M, and L stand for negligible, small, medium, and large, respectively).}
\label{tab:rq1_stats_pf_variant_model}
\resizebox{\columnwidth}{!}{\begin{tabular}{llccccc}
\toprule
PF & \success{} & DS-3.2 & GPT-OSS & QWEN-480B & QWEN-FP8 & GPT-5.2 \\
\midrule
PF1 & 1 vs. 5 & 1.00 (N) & 1.00 (N) & 1.00 (N) & 1.00 (N) & 0.45 (S) \\
 & 1 vs. 10 & 1.00 (N) & 1.00 (N) & 1.00 (N) & 1.00 (N) & 0.19 (L) \\\addlinespace[1em]

PF1 (inf.) & 1 vs. 5 & 1.00 (N) & 1.00 (N) & 1.00 (N) & 1.00 (N) & 0.45 (S) \\
 & 1 vs. 10 & 1.00 (N) & 1.00 (N) & 0.45 (S) & 1.00 (N) & 0.07 (L) \\
\midrule
PF2 & 1 vs. 5 & $\mathbf{ < 0.05 (L)}$ & $\mathbf{ < 0.05 (L)}$ & 0.12 (L) & $\mathbf{ < 0.05 (L)}$ & $\mathbf{ < 0.05 (L)}$ \\
 & 1 vs. 10 & $\mathbf{ < 0.05 (L)}$ & $\mathbf{ < 0.05 (L)}$ & 0.12 (L) & $\mathbf{ < 0.05 (L)}$ & $\mathbf{ < 0.05 (L)}$ \\\addlinespace[1em]

PF2 (inf.)  & 1 vs. 5 & $\mathbf{ < 0.05 (L)}$ & $\mathbf{ < 0.05 (L)}$ & $\mathbf{ < 0.05 (L)}$ & $\mathbf{ < 0.05 (L)}$ & $\mathbf{ < 0.05 (L)}$ \\
 & 1 vs. 10 & $\mathbf{ < 0.05 (L)}$ & $\mathbf{ < 0.05 (L)}$ & $\mathbf{ < 0.05 (L)}$ & $\mathbf{ < 0.05 (L)}$ & $\mathbf{ < 0.05 (L)}$ \\
\midrule
PF3 & 1 vs. 5 & $\mathbf{ < 0.05 (L)}$ & $\mathbf{ < 0.05 (L)}$ & 0.77 (S) & $\mathbf{ < 0.05 (L)}$ & $\mathbf{ < 0.05 (L)}$ \\
 & 1 vs. 10 & $\mathbf{ < 0.05 (L)}$ & $\mathbf{ < 0.05 (L)}$ & 0.55 (M) & $\mathbf{ < 0.05 (L)}$ & $\mathbf{ < 0.05 (L)}$ \\\addlinespace[1em]

PF3 (inf.) & 1 vs. 5 & $\mathbf{ < 0.05 (L)}$ & $\mathbf{ < 0.05 (L)}$ & 0.05 (L) & $\mathbf{ < 0.05 (L)}$ & $\mathbf{ < 0.05 (L)}$ \\
 & 1 vs. 10 & $\mathbf{ < 0.05 (L)}$ & $\mathbf{ < 0.05 (L)}$ & $\mathbf{ < 0.05 (L)}$ & $\mathbf{ < 0.05 (L)}$ & $\mathbf{ < 0.05 (L)}$ \\
\midrule
PF4 & 1 vs. 5 & 1.00 (N) & 0.45 (S) & 0.45 (S) & 0.45 (S) & 0.62 (S) \\
 & 1 vs. 10 & 1.00 (N) & 0.19 (L) & 0.45 (S) & 0.18 (L) & 0.50 (M) \\\addlinespace[1em]

PF4 (inf.) & 1 vs. 5 & 1.00 (N) & 1.00 (N) & 0.45 (S) & 1.00 (N) & 0.19 (L) \\
 & 1 vs. 10 & 1.00 (N) & 1.00 (N) & 0.45 (S) & 0.18 (L) & 0.19 (L) \\
\midrule
PF5 & 1 vs. 5 & 1.00 (N) & $\mathbf{ < 0.05 (L)}$ & 1.00 (N) & 0.45 (S) & $\mathbf{ < 0.05 (L)}$ \\
 & 1 vs. 10 & 1.00 (N) & $\mathbf{ < 0.05 (L)}$ & 0.76 (S) & 0.45 (S) & $\mathbf{ < 0.05 (L)}$ \\\addlinespace[1em]

PF5 (inf.) & 1 vs. 5 & 0.45 (S) & $\mathbf{ < 0.05 (L)}$ & 0.16 (L) & 1.00 (N) & $\mathbf{ < 0.05 (L)}$ \\
 & 1 vs. 10 & 0.45 (S) & $\mathbf{ < 0.05 (L)}$ & 0.16 (L) & 1.00 (N) & $\mathbf{ < 0.05 (L)}$ \\
\bottomrule
\end{tabular}}
\end{table}

Figure~\ref{fig:rq1}
shows violin plots with the distributions of 
\success{1}, \success{5}, and \success{10}
(\ie the probability of producing a correct solver in a single turn, 5, and 10 turns) split by \ac{pf}, feedback prompt template (\ie standard and informed) and \ac{llm} under test.
\tref{rq1_stats_pf_variant_model}
reports the p-value and effect size assessing the statistical difference of 
\success{1} vs. \success{5} and \success{1} vs. \success{10}
also split by \ac{pf}, feedback variant, and \ac{llm}.

The results in Figure~\ref{fig:rq1} show that the probability of generating a correct solver 
typically increases as turns progress. 
\pf{2} and \pf{3} exhibit the highest success rates.
Specifically, \pf{2} results in \success{1}, \success{5} and \success{10} of 29.5\%, 78.5\%, and 88\%, respectively (calculated as the mean across both feedback prompt variants and all models).
\pf{3} results in a mean \success{1}, \success{5}, and \success{10} of 25.3\%, 80.3\%, and 90.5\%, respectively.  
The same metrics are 0.5\%, 5.5\%, and 9\% for \pf{4}, and 12.8\%, 26.3\%, and 29\% for \pf{5}. 
As expected, the most complex problem family, \ie \pf{1}, results in the lowest success rates with 0\%, 0.5\%, and 2.8\%.
These differences derive from the structural complexity of the underlying tasks: \pf{2} and \pf{3} require relatively direct mappings to quantum formulations, whereas \pf{1}, \pf{4}, and \pf{5} involve more complex transformations (\eg fermion-to-qubit mappings and time evolution schemes).

As per \tref{rq1_stats_pf_variant_model}, increasing from 1 to 5 turns leads to a significant improvement in \pf{2} for all models with the informed feedback and all models but QWEN-480B with the standard feedback.
Similarly, for \pf{3}, a significant improvement is observed for all models but QWEN-480B. 
With more complex problem families, as seen in \fref{rq1}, the effectiveness decreases, also leading to less statistically significant results.
For \pf{5}, iterative feedback for 5 turns is significant with GPT models, whereas, in \pf{1} and \pf{4}, statistical significance is never achieved.
These results suggest that iterative feedback is effective in solving more superficial issues but less effective for deeper algorithmic or numerical errors, which are more likely with more complex problems.
Increasing from 5 to 10 turns is only relevant in \pf{3} with QWEN-480B where significance is observed when comparing \success{1} with \success{10} but not when comparing it with \success{5}.
 
Regarding the models under evaluation, GPT-5.2 shows the best performance. 
Finally, in this specific experimental evaluation, the difference in terms of effectiveness between the two prompt variants is never significant.
This suggests that providing the expected numerical output does not constitute a strict requirement for improvement: iterative refinement for approximately 5 turns already results in substantial gains even when only providing execution feedback.
From a practical perspective, this indicates that, in realistic scenarios where a classical reference might not be available, developers may obtain correct solutions with a limited number of iterations without relying on an oracle.

\vspace{0.15cm}
\noindent
\setlength{\fboxsep}{5pt}\setlength{\fboxrule}{1pt}\fcolorbox{gray!95}{gray!10}{\parbox{0.945\columnwidth}{\textbf{RQ1 summary.} Producing a correct solution is more likely with iterative feedback, with most gains occurring within the first five iterations. Results vary depending on the nature of the problem, with \pf{2} and \pf{3} achieving the highest success rates, while \pf{1} being the most challenging. Among models, GPT-5.2 performs best.}
    }
\vspace{0.15cm}

\subsubsection{RQ2 (Failure Causes)}

\begin{table}[t]
\centering
\caption{Failure causes taxonomy.}
\label{tab:failurecategories}
\resizebox{\columnwidth}{!}{
\begin{tabular}{llp{3.5cm}p{5.5cm}}
\toprule
ID & Cause & Explanation & Keywords \\
\midrule
NumErr & Numerical Error & Code runs but produces wrong result. & - \\\addlinespace[1em]
Timeout & Timeout & Code execution halted due to timeout. & - \\\addlinespace[1em]
API & Library API Misuse & Incorrect API usage. & unexpected keyword argument, got an unexpected keyword, missing required positional argument, TypeError, AttributeError, Qiskit error \\\addlinespace[1em]
Deps & Version / Dependency Issue & Missing/incompatible deps. & cannot import name, cannot import, no module named, ImportError, ModuleNotFoundError \\\addlinespace[1em]
Type & Type / Data Structure Error & Invalid types/structures. & unsupported operand type, invalid value, KeyError, ValueError \\\addlinespace[1em]
Gen & Code Generation Error & Invalid/generated code. & name is not defined, invalid syntax, NameError, SyntaxError, IndentationError \\
\bottomrule
\end{tabular}}
\end{table}

\begin{figure}[t]
\centering
    \includegraphics[width=\columnwidth]{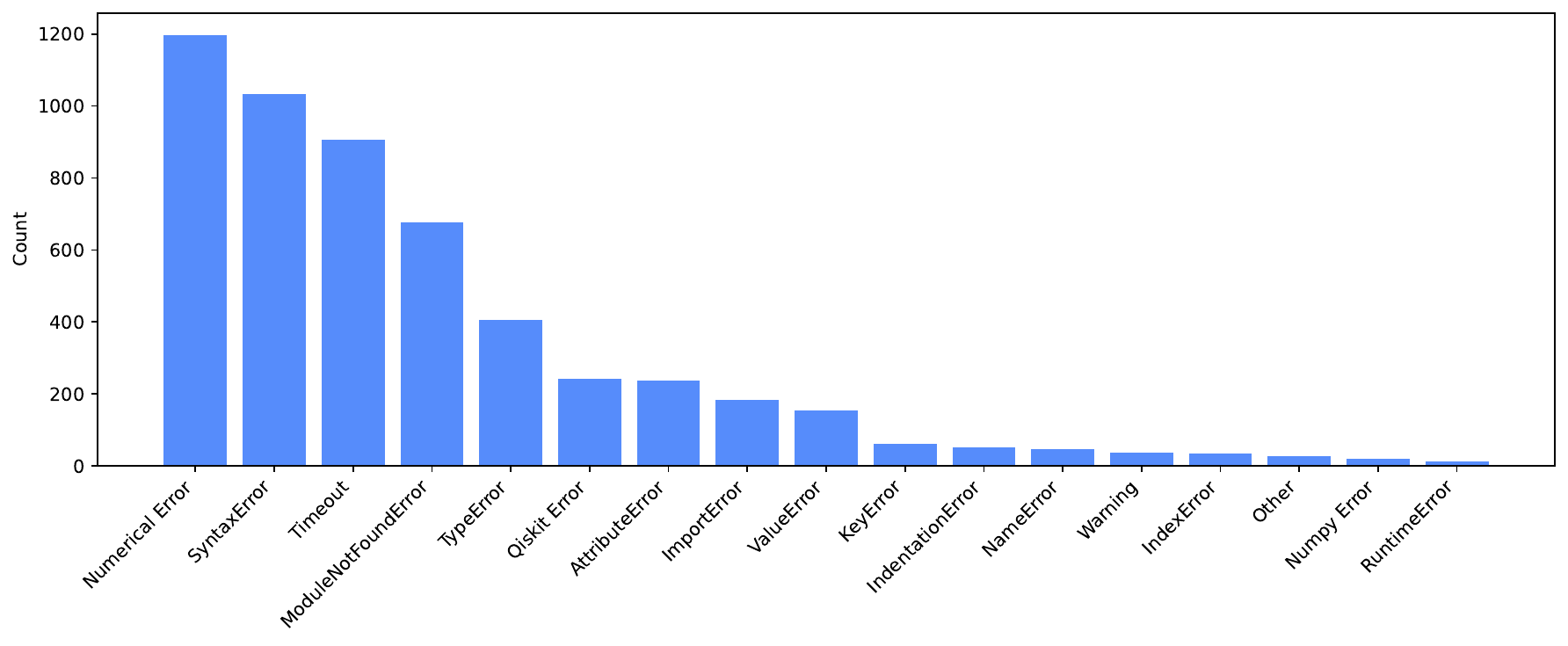}
    \caption{Histogram of errors (\ie exceptions, warnings, and output errors).}
    \label{fig:hist}
\end{figure}

\begin{table}[t]
\centering
\caption{Failure cause percentage by \ac{pf} and model (the most significant category for each pair is in red). ``N.Fail.'' is the number of failures out of the 800 attempts.}
\label{tab:errordistr}
\resizebox{1\columnwidth}{!}{
\begin{tabular}{llllllllll}
\toprule
PF & Model & N.Fail. & NumErr & Timeout & API & Deps & Type & Gen & Other \\
\midrule

PF1 & DS-3.2 & 800 & 0.0\% & 22.6\% & 9.4\% & 11.5\% & 2.1\% & \textbf{\textcolor{red}{50.7\%}} & 3.7\% \\
PF1 & GPT-OSS & 800 & 0.5\% & 15.1\% & \textbf{\textcolor{red}{38.4\%}} & 33.6\% & 1.0\% & 10.0\% & 1.4\% \\
PF1 & QWEN-480B & 792 & 0.4\% & 1.4\% & 22.1\% & \textbf{\textcolor{red}{69.2\%}} & 1.0\% & 1.0\% & 4.9\% \\
PF1 & QWEN-FP8 & 800 & 0.0\% & 5.9\% & 13.1\% & 36.8\% & 0.0\% & \textbf{\textcolor{red}{40.2\%}} & 4.0\% \\
PF1 & GPT-5.2 & 762 & 0.1\% & \textbf{\textcolor{red}{74.5\%}} & 13.3\% & 0.5\% & 8.3\% & 0.0\% & 3.3\% \\[.6em]

PF2 & DS-3.2 & 231 & 13.9\% & 21.6\% & 10.4\% & 0.4\% & 4.3\% & \textbf{\textcolor{red}{43.7\%}} & 5.6\% \\
PF2 & GPT-OSS & 388 & 3.6\% & 1.0\% & \textbf{\textcolor{red}{82.0\%}} & 1.3\% & 0.3\% & 0.0\% & 11.9\% \\
PF2 & QWEN-480B & 72 & \textbf{\textcolor{red}{50.0\%}} & 20.8\% & 11.1\% & 2.8\% & 15.3\% & 0.0\% & 0.0\% \\
PF2 & QWEN-FP8 & 370 & 17.0\% & 15.1\% & \textbf{\textcolor{red}{34.6\%}} & 7.3\% & 23.2\% & 0.5\% & 2.2\% \\
PF2 & GPT-5.2 & 45 & \textbf{\textcolor{red}{37.8\%}} & 2.2\% & 22.2\% & 0.0\% & 22.2\% & 2.2\% & 13.3\% \\[.6em]

PF3 & DS-3.2 & 359 & 21.2\% & 8.6\% & 0.0\% & 15.6\% & 0.0\% & \textbf{\textcolor{red}{52.1\%}} & 2.5\% \\
PF3 & GPT-OSS & 152 & 36.8\% & 0.0\% & 0.0\% & \textbf{\textcolor{red}{52.6\%}} & 0.0\% & 0.0\% & 10.5\% \\
PF3 & QWEN-480B & 274 & \textbf{\textcolor{red}{65.7\%}} & 26.6\% & 0.0\% & 6.6\% & 0.7\% & 0.4\% & 0.0\% \\
PF3 & QWEN-FP8 & 272 & 42.6\% & 3.7\% & 0.4\% & \textbf{\textcolor{red}{51.8\%}} & 0.0\% & 0.4\% & 1.1\% \\
PF3 & GPT-5.2 & 42 & \textbf{\textcolor{red}{52.4\%}} & 16.7\% & 0.0\% & 26.2\% & 0.0\% & 2.4\% & 2.4\% \\[.6em]

PF4 & DS-3.2 & 800 & 0.4\% & 3.6\% & 0.8\% & 0.2\% & 0.1\% & \textbf{\textcolor{red}{93.1\%}} & 1.8\% \\
PF4 & GPT-OSS & 784 & 11.5\% & \textbf{\textcolor{red}{61.2\%}} & 7.3\% & 0.8\% & 3.3\% & 11.0\% & 5.0\% \\
PF4 & QWEN-480B & 742 & 32.9\% & 2.8\% & \textbf{\textcolor{red}{48.7\%}} & 5.0\% & 5.1\% & 1.5\% & 4.0\% \\
PF4 & QWEN-FP8 & 786 & 7.8\% & 15.0\% & \textbf{\textcolor{red}{40.7\%}} & 14.2\% & 4.3\% & 7.6\% & 10.3\% \\
PF4 & GPT-5.2 & 670 & \textbf{\textcolor{red}{71.3\%}} & 16.0\% & 1.8\% & 0.0\% & 2.1\% & 0.0\% & 8.8\% \\[.6em]

PF5 & DS-3.2 & 793 & 2.8\% & 3.3\% & 0.9\% & 0.1\% & 0.4\% & \textbf{\textcolor{red}{91.6\%}} & 1.0\% \\
PF5 & GPT-5.2 & 43 & \textbf{\textcolor{red}{90.7\%}} & 0.0\% & 4.7\% & 0.0\% & 0.0\% & 0.0\% & 4.7\% \\
PF5 & GPT-OSS & 670 & \textbf{\textcolor{red}{53.0\%}} & 14.8\% & 11.2\% & 0.1\% & 0.0\% & 13.6\% & 7.3\% \\
PF5 & QWEN-480B & 712 & \textbf{\textcolor{red}{87.2\%}} & 0.0\% & 5.3\% & 0.0\% & 6.0\% & 0.0\% & 1.4\% \\
PF5 & QWEN-FP8 & 791 & \textbf{\textcolor{red}{55.5\%}} & 6.1\% & 28.7\% & 0.1\% & 0.4\% & 5.6\% & 3.7\% \\
\bottomrule
\end{tabular}
}
\end{table}

For this analysis, we distinguish between \emph{failures}, \emph{errors}, and \emph{failure causes} to avoid ambiguity.
A \emph{failure} denotes a run in which the generated solver does not meet the correctness criterion.
An \emph{error} is a low-level signal observed during execution, such as an exception, warning, timeout, or numerical mismatch.
A \emph{failure cause} is a higher-level abstraction obtained by grouping errors to enable systematic analysis.
To elicit and analyze failure causes, we adopt a classification pipeline built on top of \approachName{} execution traces.
For each run recording a failure, we apply a rule-based classifier that inspects error messages, exceptions, warnings, and numerical outputs.
\fref{hist} shows the most common exceptions, including numerical errors (\ie the generated code produces a result that differs from the ground truth), warnings and timeouts.

Low-level traces are mapped to higher-level failure causes, namely incorrect output, timeout, API misuse, dependency issues, invalid types/structures, and code generation errors (also summarized in \tref{failurecategories}).
After classification, failures are aggregated across problem families and models. 
Finally, we compute the percentage distributions on the failure causes, reported in \tref{errordistr}, providing a comparative view of the dominant error modes exhibited by different \acp{llm} and problem families.
In \tref{errordistr}, the $25\%$ least common errors are grouped under the ``Other'' column.

The results highlight different trends for different models and \acp{pf}.
Concerning models, DS-3.2 consistently shows code generation as the prevalent cause of failure, mostly due to the model failing to respond without unformatted natural language explanations (despite it being explicitly enforced in the prompt).
GPT-5.2 shows the best performance, as it more frequently generates running code with the prevalent causes of failure being timeout for \pf{1} and numerical errors in all other cases.
GPT-OSS, QWEN-480B, and QWEN-FP8 exhibit a more mixed behavior, with API misuse and version/dependency issues being prevalent in some cases.
Overall, these results suggest a progression in failure causes: as model capability and size increase, errors shift from syntactic and integration issues to semantic and numerical inaccuracies.

Concerning problem families, \pf{5} shows a clear prevalence of numerical errors. 
This is aligned with the nature of quantum chemistry problems, where implementations are more standardized and less prone to syntactic errors but correctness strictly depends on precise parameter handling.
In the other cases, the trends are more mixed, with API misuses (thus, indicating lack of dependency knowledge on the \acp{llm} side) being more prevalent in \pf{2} and \pf{4}. 
This reflects the reliance on specific quantum software stacks and non-trivial library interactions, which remain challenging for current models.

These results have practical implications on the design of \ac{llm}-based workflows, suggesting that different mitigation strategies might be required depending on the dominant failure mode.
For example, stronger prompt constraints and validation are in order when code generation errors are prevalent, while providing additional domain-specific guidance is necessary under significant residual numerical inaccuracies.

\vspace{0.15cm}
\noindent
\setlength{\fboxsep}{5pt}\setlength{\fboxrule}{1pt}\fcolorbox{gray!95}{gray!10}{\parbox{0.945\columnwidth}{\textbf{RQ2 summary.} Failure modes vary across models and problem families: DS-3.2 is dominated by code generation errors, while GPT-5.2 mainly fails due to timeouts and numerical inaccuracies. Other models show mixed behavior, with API misuse and dependency issues frequently emerging, especially in \pf{2} and \pf{4}, whereas \pf{5} is largely dominated by numerical errors.}
    }
\vspace{0.15cm}

\subsubsection{RQ3 (Cost)}

\begin{figure}[t]
    \centering
    \begin{subfigure}{\columnwidth}
        \centering
        \includegraphics[width=.35\columnwidth]{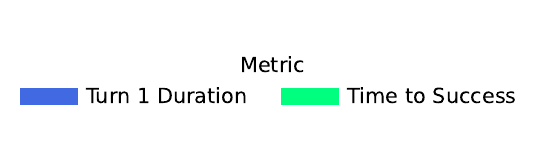}
    \end{subfigure}
    \begin{subfigure}{\columnwidth}
        \centering
        \includegraphics[width=\columnwidth]{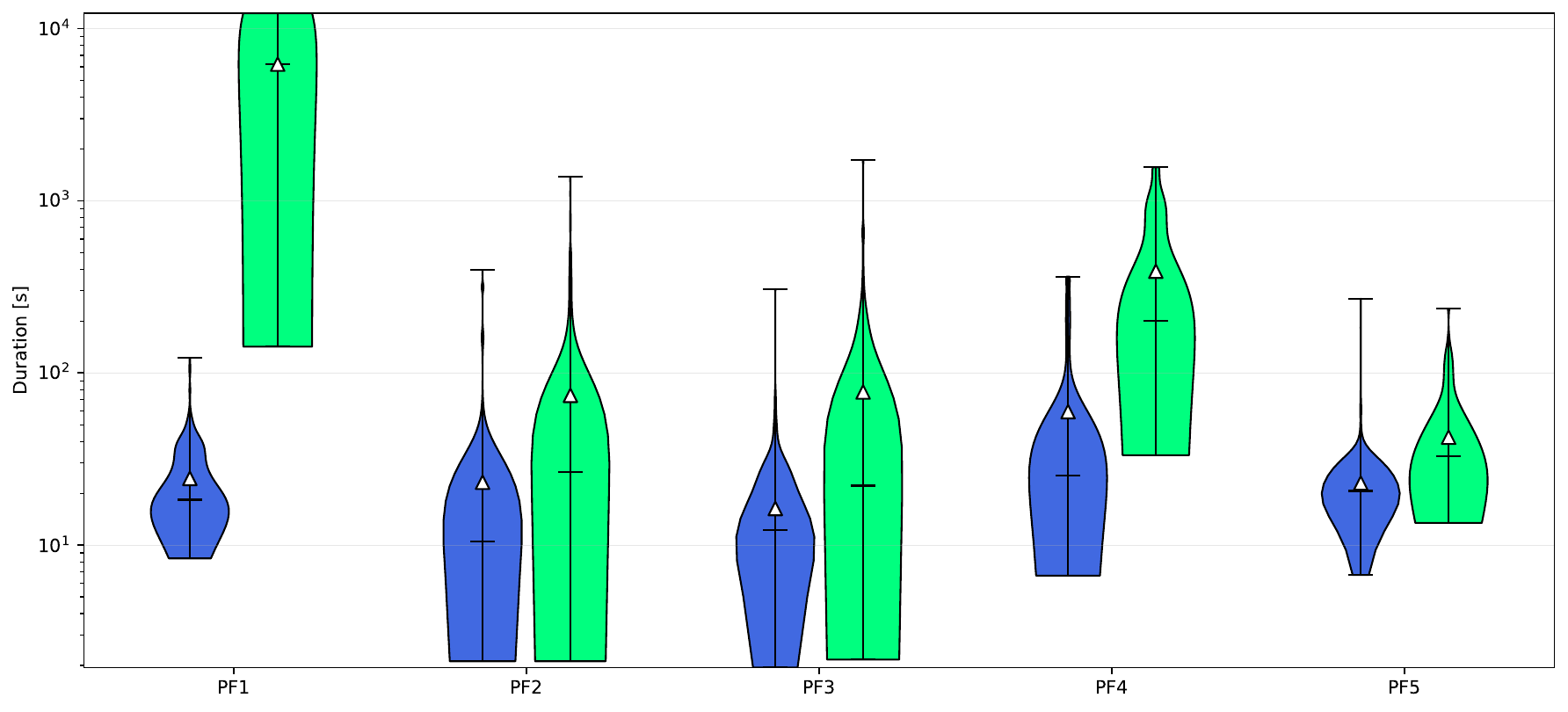}
        \caption{Test duration per \ac{pf}.}
        \label{fig:timepf}
    \end{subfigure}
    \begin{subfigure}{\columnwidth}
        \centering
        \includegraphics[width=\columnwidth]{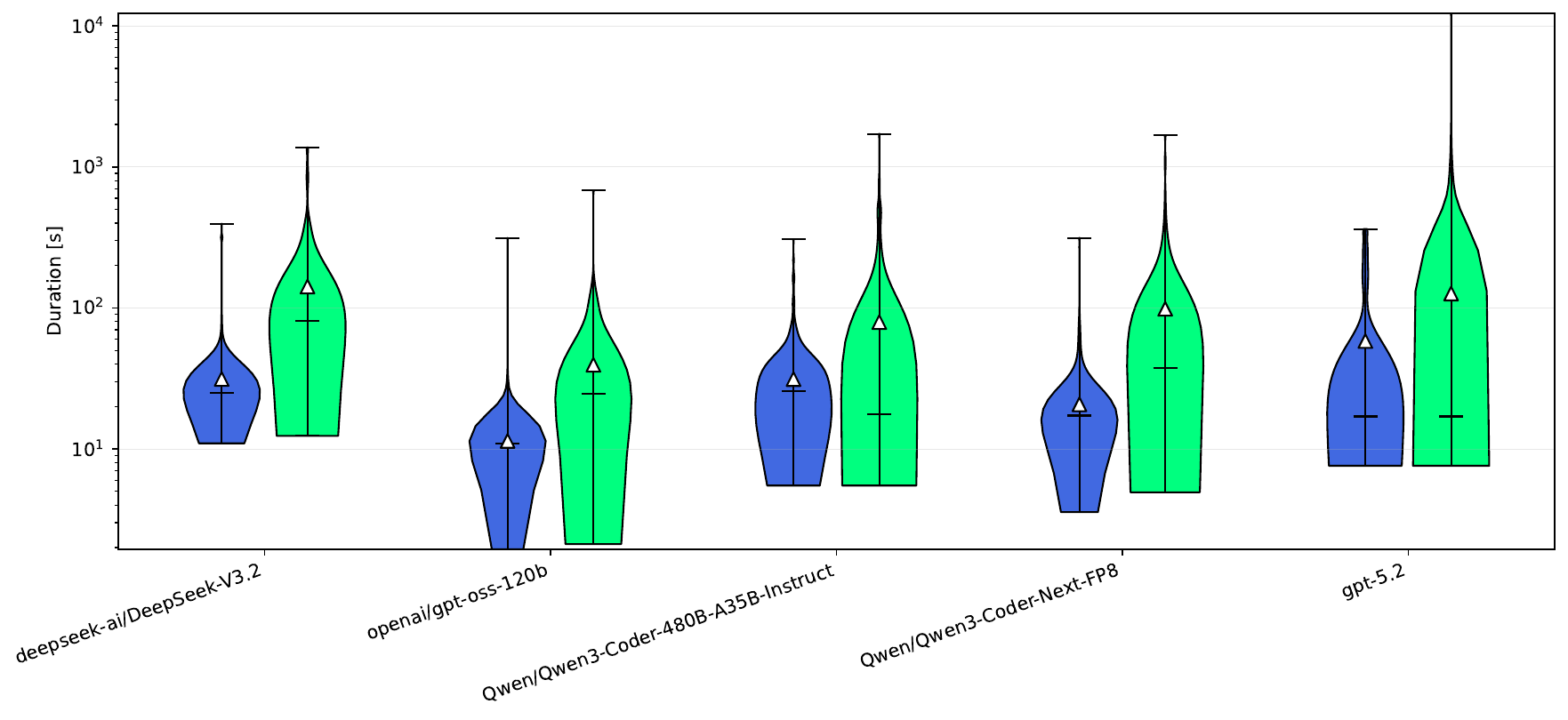}
        \caption{Test duration per model.}
        \label{fig:timemodel}
    \end{subfigure}
\caption{Test duration---in log-scale---by \ac{pf} and model. The plots compare the duration of the first turn (the baseline, in blue) with the cumulative time to the turn where success is recorded (in green).}
\label{fig:rq3}
\end{figure}

Beyond effectiveness, the practical viability of \approachName depends on the computational cost required to obtain correct solutions.
To assess the time overhead introduced by the iterative \approachName{} methodology, we compare the duration of turn 1 with the time required to reach the first successful solution.
We perform this analysis across all problem families and models, using execution time measured in seconds.
The distributions of execution times are reported in \fref{rq3} by model and problem family.

The distributions show that, while turn 1 executions are relatively short and concentrated, the time to first success exhibits a much wider spread and longer tails, with some runs requiring orders of magnitude more time.
This effect is particularly evident in more complex \acp{pf} such as \pf{1} and \pf{4}, where iterative refinement leads to longer executions and occasional extreme outliers.
This highlights a clear trade-off: while iterative refinement improves correctness, it comes at the cost of increased and variable execution time.

We statistically compare turn 1 duration and the time to first success using the Mann–Whitney U test and the Vargha–Delaney effect size measure. 
The results show that the differences are statistically significant (${p < 0.05}$) with large effect sizes in all problem families.
This indicates that the time required to obtain a correct solution is consistently and substantially higher than the cost of a single generation attempt.

Taking into account the statistical metrics between the models, the cost of additional iterations is significant (${p=0.05}$ with a large effect size) for DS-3.2, GPT-OSS, and QWEN-FP8, while it is not for QWEN-480B and GPT-5.2 (${p=0.41}$ and ${p=0.11}$, respectively). 
In combination with the RQ1 results, this confirms that when QWEN-480B passes the test, it is most likely to do so at an early turn. With GPT-5.2, this interestingly shows that iterating for 5 turns leads to a significant effectiveness increase in light of a more contained time overhead.

From a practical perspective, this suggests that \approachName is most suitable in scenarios where correctness is critical and additional computational cost is acceptable but its applicability might be limited in time-constrained settings.

\vspace{0.15cm}
\noindent
\setlength{\fboxsep}{5pt}\setlength{\fboxrule}{1pt}\fcolorbox{gray!95}{gray!10}{\parbox{0.945\columnwidth}{\textbf{RQ3 summary.} Achieving a correct solution requires significantly more time than a single generation attempt across all problem families (large effect sizes), with substantial variability and long-tail behavior, especially for more complex tasks. }
    }
\vspace{0.15cm}

\subsection{Threats to Validity}
\label{sec:threats}

Our results rely on the correctness of the oracle used for verification. Any inaccuracies in the oracle implementation or tolerance thresholds affect the pass/fail classification. To mitigate this risk, we rely on established expert-validated classical solvers with conservative tolerances. 
Additionally, failure classification is based on a rule-based analysis of execution traces, which can introduce misclassification errors (\eg ambiguous error messages or overlapping symptoms). However, the taxonomy is grounded in explicit signals (exceptions, timeouts, numerical mismatches) to reduce subjectivity.

The evaluation focuses on five \acp{pf} and a specific set of parameter configurations, which, while representative of different domains (many-body physics, combinatorial optimization, quantum chemistry, and lattice gauge theory), may not generalize to larger-scale problems, alternative quantum algorithms, or domains lacking tractable classical references. 
Moreover, we evaluate a limited set of \acp{llm} at specific versions and configurations: results may differ for newer models or alternative prompting strategies.

The coder and feedback prompt templates used in the experiments were crafted to reflect realistic usage scenarios and are consistent across models to ensure a fair comparison. However, there may exist alternative prompt formulations or different prompt engineering strategies leading to different performance outcomes.
The controlled execution environment (\ie the fixed library versions and resource constraints) may also affect generalizability, since different environments might induce different failure causes and success rates.

We use non-parametric statistical tests (Mann-Whitney U) and effect size measures (Vargha-Delaney) following established guidelines to assess significance and magnitude. To reduce the impact of stochastic variability in \ac{llm} outputs, each experiment is repeated multiple times. 

\subsection{Discussion}
\label{sec:discu}

Our results highlight the current limitations of \acp{llm} in generating correct scientific code for computational scientific tasks.

First, the effectiveness results (RQ1) show that iterative refinement significantly improves success rates, with most gains occurring within the first 5 iterations.
This indicates that \acp{llm} benefit from execution feedback, particularly for resolving syntactic and integration issues. However, the diminishing returns observed beyond 5 turns indicate that additional iterations are less effective in addressing deeper correctness problems.

Second, the failure analysis (RQ2) reveals that errors are not uniformly distributed but depend strongly on both the model and the problem family. Less capable models are primarily affected by code generation errors and API misuse, indicating difficulties in producing executable and syntactically valid code. In contrast, stronger models tend to fail due to numerical inaccuracies, highlighting a shift from execution-related issues to semantic correctness challenges. This suggests that as model capability increases, the main bottleneck moves from generating runnable code to producing scientifically correct implementations.

Third, the cost analysis (RQ3) exposes a clear trade-off between correctness and efficiency. While iterative refinement improves success rates, it introduces a substantial time overhead, with time to first success significantly exceeding the cost of a single generation attempt and exhibiting long-tail behavior. This overhead is particularly pronounced for more complex problem families, where failures require multiple iterations to resolve. Nevertheless, for high-performing models such as GPT-5.2, the additional cost is relatively more contained, suggesting that stronger models can achieve better efficiency-effectiveness trade-offs.

In general, these findings point to a capability limit for current \acp{llm} in scientific code generation.
Although they can effectively help produce executable implementations and benefit from iterative feedback, they still struggle with domain-specific correctness and numerical reliability. This has important implications for the design of \ac{llm}-based systems in scientific computing: verification mechanisms, domain-specific guidance, and improved integration with numerical libraries are essential to ensure reliable results.

\section{Conclusions}
\label{sec:concl}

This work has evaluated the ability of \acp{llm} to generate correct quantum solver code for computational scientific problems using \approachName, an execution-based methodology combining code generation, execution, and oracle-driven verification.
The results show that iterative refinement significantly improves success rates, and most gains occur within the first few turns. However, this comes at a non-negligible cost, as reaching a correct solution requires substantially more time than a single generation attempt.
The failure category analysis reveals some error patterns: weaker models primarily fail due to code generation and API issues, while stronger models more often produce executable but numerically incorrect solutions.

In general, our findings highlight a key limitation of current \acp{llm} in scientific computing: correctness cannot be inferred from the execution alone, and numerical accuracy remains a major challenge.
Future work includes improving feedback and prompting strategies to guide models toward correct solutions more efficiently, integrating domain-specific knowledge or verification mechanisms, and extending the evaluation to more complex problems and emerging \ac{llm} architectures.

\bibliographystyle{IEEEtran}

\end{document}